\documentclass{article}
\usepackage{ijcai22}

\usepackage{times}
\usepackage{soul}
\usepackage{url}
\usepackage[hidelinks]{hyperref}
\usepackage[utf8]{inputenc}
\usepackage[small]{caption}
\usepackage{graphicx}
\usepackage{amsmath}
\usepackage{amsthm}
\usepackage{booktabs}
\usepackage{algorithm}
\usepackage{algorithmic}
\usepackage{multirow}
\usepackage{makecell}

\title{BagFormer: Better Cross-Modal Retrieval via bag-wise interaction}

\author{
Haowen Hou$^1$
\and
Xiaopeng Yan$^1$\and
Yigeng Zhang$^2$\and
Fengzong Lian$^1$\And
Zhanhui Kang$^1$
\affiliations
$^1$Tencent Inc., Shenzhen, China\\
$^2$University of Houston,  Houston, USA
\emails
haowen.hou@u.nus.edu,
chopinyan@tencent.com,
yzhang168@uh.edu
}

\begin{document}

\maketitle

\begin{abstract}
In the field of cross-modal retrieval, single encoder models tend to perform better than dual encoder models, but they suffer from high latency and low throughput.
In this paper, we present a dual encoder model called BagFormer that utilizes a cross modal interaction mechanism to improve recall performance without sacrificing latency and throughput.
BagFormer achieves this through the use of bag-wise interactions, which allow for the transformation of text to a more appropriate granularity and the incorporation of entity knowledge into the model. 
Our experiments demonstrate that BagFormer is able to achieve results comparable to state-of-the-art single encoder models in cross-modal retrieval tasks, while also offering efficient training and inference with 20.72 times lower latency and 25.74 times higher throughput.
\end{abstract}

\section{Introduction}

\begin{figure*}[t]
\centering
\includegraphics[scale=0.6]{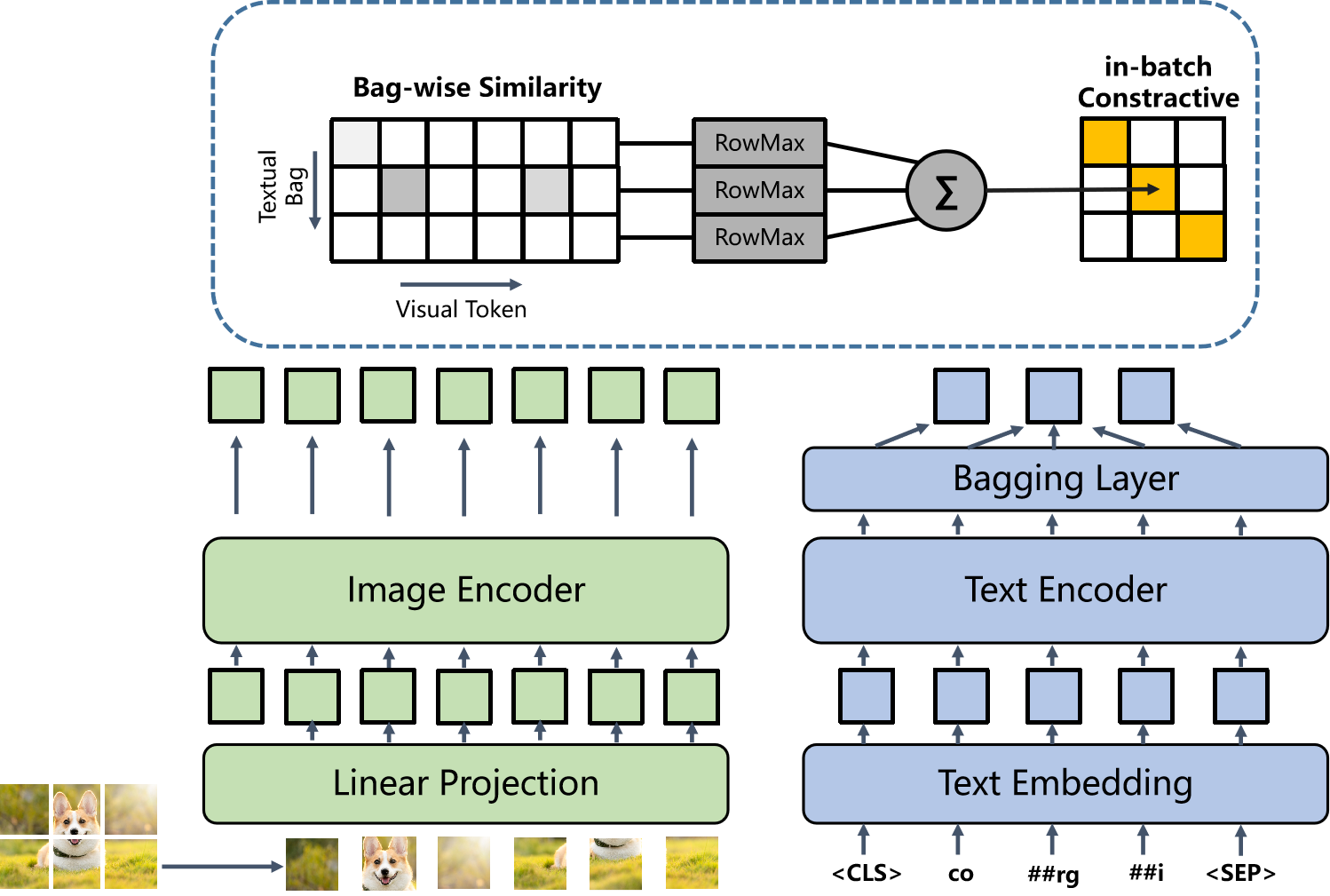}
\caption{The architecture of BagFormer network. A Dual-encoder architecture with Transformer-based image encoder and text encoder. Textual tokens in a 'bag' are aggregated to textual bags by the bagging layer. Then the representations of textual bags and visual tokens are linearly projected into the multi-modal joint space. We propose a novel bag-wise interaction method based on bag-wise maximum similarity between visual tokens and textual bags.}
\end{figure*}

Cross-modal retrieval\cite{faghri2017vse++,zeng2020deep} is used for implementing a retrieval task across different modalities(such as image-text or text-image) and is considered one of the most important multimodal understanding tasks\cite{lee2018stacked}. The performance of this retrieval task is largely improved by better visual representation and detailed image-text alignment \cite{cheng2022vista}. Recently, cross-modal retrieval has recently been greatly advanced by the development of Vision-Language Pre-Training (VLP), which are capable of learning visual and textual representations from millions of images and texts collected from the Internet and have superior zero-shot ability and robustness\cite{kim2021vilt,lu2019vilbert,li2020unicoder}.

There are mainly two types of cross-modal retrieval architectures: single-encoder and dual-encoder. Single-encoder models like Visual-BERT\cite{li2019visualbert} and ALBEF\cite{li2021align} directly feed visual features and textual embeddings to the transformer-based model. Dual-encoder models such as CLIP\cite{radford2021learning} and ALIGN\cite{jia2021scaling} have separate encoders for vision and language. In general, dual encoder models are faster but less effective than single encoder models. 
On the other hand, single-encoder models outperform dual-encoder models in recall metric, but they are too slow\cite{khattab2020colbert}.
Generally speaking, dual-encoder models lack modal interaction and are therefore less effective than single-encoder models\cite{khattab2020colbert}. 
In this paper, we propose BagFormer, a dual-encoder model with bag-wise interaction mechanism for cross-modal retrieval, which greatly reduces the recall metric gap between the single-encoder models and the dual-encoder models. 

Cross-modal interaction and cross-modal alignment of image and text are the key to succeed in cross-modal retrieval\cite{li2021align,yao2021filip}.
In order to achieve fine-grained cross-modal interaction and alignment, previous approaches have mainly based on a two kinds of methods.
In a series of works, pre-trained object detectors\cite{lee2018stacked,anderson2018bottom} have been used to extract ROI features from images and combine them with paired text. This approach can be a bit more challenging since a large number of ROI features must be computed and stored in advance\cite{li2021align}. Furthermore, the zero-shot ability of these approaches is typically limited by the number of predefined classes, and their performance is also governed by the quality of the detector\cite{li2021align}.
Another series of work introduce interaction between modalities. Single encoder models use token-wise or patch-wise representations and achieves fine-grained interactions via cross-attention or self-attention\cite{cheng2022vista}. However, these methods are typically less efficient in terms of both training and inference. Particularly during training, cross-attention needs to be performed in an encoder-decoder structure, whereas self-attention becomes increasingly complex as it is concatenated with both image and text for extended sequences. For inference, an image-text pair must be combined in order to compute cross-attention or self-attention. Therefore, the single encoder models cannot be pre-computed off-line like dual encoder models, such as CLIP\cite{radford2021learning} and ALIGN\cite{jia2021scaling}. 
Cross-attention or self-attention used in single encoder models is too expensive in computation, so ColBERT\cite{khattab2020colbert,santhanam2021colbertv2} proposes to introduce a late interaction mechanism into the dual encoder models to achieve efficient and effective passage retrieval. FILIP\cite{yao2021filip} introduces the idea of late interaction into the multi-modal domain.

Our paper aims to optimize the cross-modal interaction so that the dual encoder model's performance is close to the single encoder model's with significantly lower latency and higher throughput. 
To this end, we propose BagFormer, a dual encoder model with bag-wise interaction to achieve better cross modal interaction and alignment. 
The BagFormer adopts the bag-wise interaction mechanism, which can transform text to appropriate granularity and implicitly introduce entity knowledge into the model. 
More specifically, BagFormer uses a bag-wise maximum similarity between visual and textual tokens to guide the contrastive objective. 
In this way, BagFormer successfully leverages the fine-grained expressiveness among image patches and textual words while simultaneously gaining the ability to pre-compute image and text representations off-line.

Several experiments have demonstrated that BagFormer's learning of bag representations allows it to perform at the top of its class. On image-text retrieval, BagFormer performs almost as well as the state of the art single encoder model, but the latency is reduced by 20.72 times, and the throughput is increased by 25.74 times. In addition, BagFormer outperforms other dual encoder models that are pre-trained on larger datasets. 
Furthermore, visualizations of word patch alignment demonstrate BagFormer is able to learn meaningful fine-grained features with promising localization capabilities. 

\section{Related Work}

\textbf{Vision language pre-training}. As a mainstream paradigm in multi-modal understanding, vision language pre-training has significantly increased performance on various vision and language tasks. The majority of these approaches use transformer-based architectures, which can be categorized as single-encoder and dual-encoder. With single encoder architectures, such as ALBEF\cite{li2021align}, multi-modal transformers can simultaneously process images and text for high-performance interactions. The computation cost of these approaches, however, is still impractical for large-scale cross-modal retrieval tasks. In contrast, dual encoder architectures, such as CLIP\cite{radford2021learning} and ALIGN\cite{jia2021scaling}, encode images and text separately, making it possible to calculate image-text similarities in linear time. Although the million-scale image-text contrastive pre-training greatly improves the dual encoder architecture, a performance gap still exists between it and single encoder architecture. By contrast, BagFormer incorporates bag-wise interaction into dual encoder architecture, reducing performance gap while increasing inference speed.

\textbf{Dense retrieval}. Dense retrieval involves using dense vectors to represent queries and documents and using techniques such as inner product or cosine similarity to measure their similarity\cite{wang2022neural}. This research area has recently been advanced by the use of pre-trained language models like BERT\cite{devlin2018bert} and RoBERTa\cite{liu2019roberta}, which can generate dense representations for queries and documents. To efficiently search for relevant documents, approximate nearest neighbor algorithms such as k-dimensional trees\cite{bentley1975multidimensional}, locality-sensitive hashing\cite{datar2004locality}, and graph-based indexes\cite{malkov2018efficient} can be utilized to retrieve documents in sublinear time. In addition, ColBERT\cite{khattab2020colbert} and ColBERTv2\cite{santhanam2021colbertv2} developed a late interaction paradigm that uses a scalable "MaxSim" operator for query-document interaction to boost search quality.  

\textbf{Cross-modal retrieval}. Cross-modal retrieval is the task of retrieving relevant images or text descriptions given a text or image query. In recent years, the visual representation for cross-modal retrieval has been improved through the use of techniques such as grid-based CNNs\cite{faghri2017vse++}, pre-trained object detectors\cite{anderson2018bottom} and vision-language pre-training models\cite{radford2021learning}. At the same time, there have been advances in methods for aligning images and text, including the use of attention mechanisms\cite{li2021align}, iterative matching\cite{chen2020imram}, graph-based relationship reasoning\cite{liu2020graph} and late interaction mechanism\cite{khattab2020colbert}. FILIP\cite{yao2021filip} borrow this idea of late interaction mechanism from ColBERT\cite{khattab2020colbert},  and develop a token-wise interaction for better cross-modal retrieval. BagFormer takes this approach a step further by introducing the idea of bag-wise interactions.

\section{BagFormer}

\subsection{Model Architecture}
\textbf{Model Architecture}. BagFormer integrates an image encoder based on a 12-layer visual transformer (ViT-B/16)\cite{dosovitskiy2020image} with weights pre-trained on ImageNet-1k\cite{touvron2021training} and a text encoder based on a 6-layer transformer\cite{vaswani2017attention} initialised with the first 6 layers of the BERT-base\cite{devlin2018bert} model. The input image and text are converted into respective sequences of visual embeddings \{\(v_{cls}, v_{1}, ..., v_{n}\)\} and token embeddings \{\(t_{cls}, t_{1}, ..., t_{m}\)\}. These embeddings are then projected into a common multimodal space via linear transformation, and subsequently normalised using L2-normalisation. 
Finally, the token embeddings are aggregated to form bag embeddings via the bagging layer, and the bag-wise similarity is then calculated by the late interaction of the visual embeddings and bag embeddings.

\subsection{Pre-training Objectives}
The two objectives of BagFormer's pre-training are image-text contrastive learning (ITC) and bag-wise contrastive learning (BWC).

\textbf{Image-Text Contrastive Learning} aims to align the global  representations of image and text. In CLIP\cite{radford2021learning}, [CLS] token of image patches and text tokens are used to compute the global feature, and the similarity is computed via the dot product of the global feature. More specifically, the global similarity between the image and the text is computed as follows:
\begin{equation}
s(I, T) = s(T, I) = g_v(v_{cls})^Tg_t(t_{cls})
\end{equation}
where \(g_v(v_{cls})\) denotes the embedding of the [CLS] token of the image and \(g_v(v_{cls})\) denotes the embedding of the [CLS] token of the text. \(g_v\) is the visual projection head and \(g_t\) is the textual projection head.

For each image and text, we calculate the image-to-text and text-to-image contrastive loss as
\begin{equation}
L_{i2t} = -\frac{1}{bs}log\frac{exp(s(I,T)/\tau)}{\sum_{n=1}^{bs}exp(s(I,T)/\tau)}
\end{equation}
\begin{equation}
L_{t2i} = -\frac{1}{bs}log\frac{exp(s(T,I)/\tau)}{\sum_{n=1}^{bs}exp(s(T,I)/\tau)}
\end{equation}
where \(\tau\) is a learnable temperature parameter and \(bs\) is the batch size. The total image-text contrastive loss of one training batch is then computed as follows: 
\begin{equation}
L_{itc} = \frac{1}{2}(L_{t2i} + L_{i2t})
\end{equation}

\textbf{Bag-wise Contrastive Learning}. Bag-wise Contrastive Learning is based on the bag-wise similarity, which is computed using a fine-grained interaction between image patches and text bags. A text bag is calculated by summing the embeddings of text tokens in a 'bag'. The bagging layer transforms the last-layer token embeddings \{\(t_{cls}, t_{1}, ..., t_{m}\)\} into a sequence of bag embeddings \{\(b_{cls}, b_{1}, ..., b_{k}\)\}. Each visual token \(g_v(v_{i})\) computes similarities with all non-padded textual bags, followed by selecting the maximum value of these similarities as the activation of that visual token. Finally, we average the activation of all visual tokens as the bag-wise image-to-text similarity.

\begin{equation}
s(I, T)_{bag} = \frac{1}{n}{\sum_{i=1}^{n}\max(g_v(v_{i})^Tg_t(b_{1}...b_{k}))}
\end{equation}

where \(n\) is the sequence length of image patches and \(k\) is the sequence length of text bags.  The bag-wise text-to-image similarity can be computed in the same way. 

\subsection{Bagging Layer}

\begin{figure}[htbp]
\centering
\includegraphics[scale=0.4]{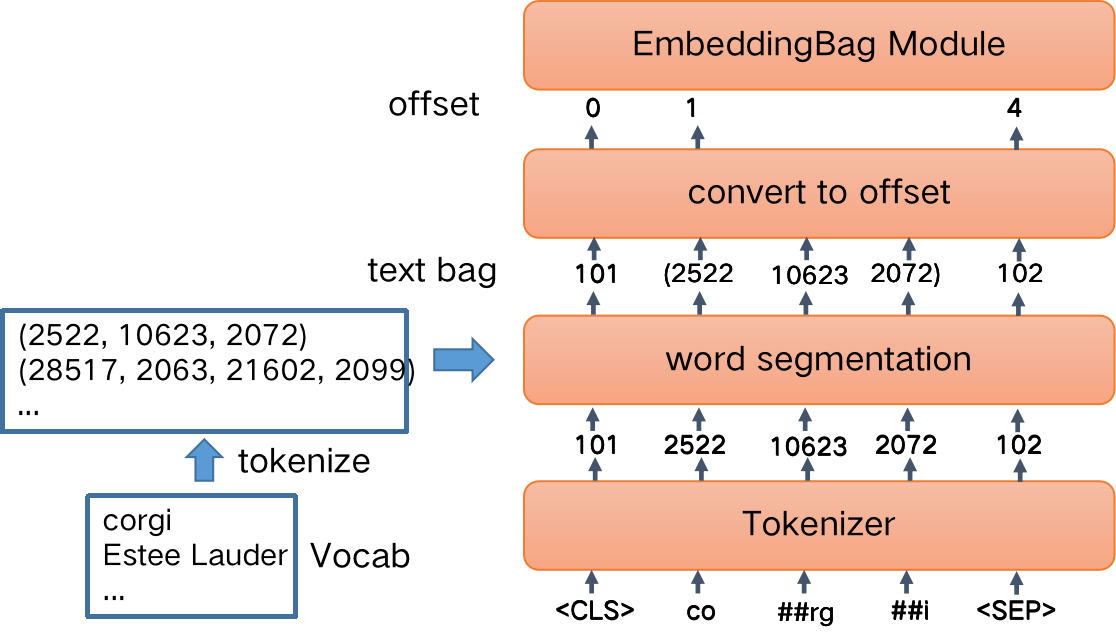}
\caption{An illustration of the bagging layer.}
\end{figure}

The purpose of bagging layer is to adjust the token granularity through by aggregating tokens into a bag, which can be a word, an entity or a phrase.
At the same time, the bagging layer is able to implicitly add a priori knowledge from the vocabulary to the model for better alignment for text and images. 
The vocabulary is a very useful priori knowledge that can be plugged into the model to enhance it. Vocabulary can be gathered manually or through data mining algorithms, such as AutoPhrase\cite{shang2018automated}.

The workflow of the bagging layer can be divided into 3 steps. The first step is to train a bagging helper, which will store the index of tokenized word/entity/phrase. In the second step, with the help of word segmentation algorithm, the input token index will be bagged in a 'bag'. In the last step, the index bag will be converted to offset, which will be fed into EmbeddingBag module. EmbeddingBag module is a high efficient implement for summing of ‘bags’ of embeddings, without instantiating the intermediate embeddings. Finally, we have a sequence of bag embedding as the layer's output.

\subsection{Early bagging vs. Late bagging}

\begin{figure}[htbp]
\centering
\includegraphics[scale=0.35]{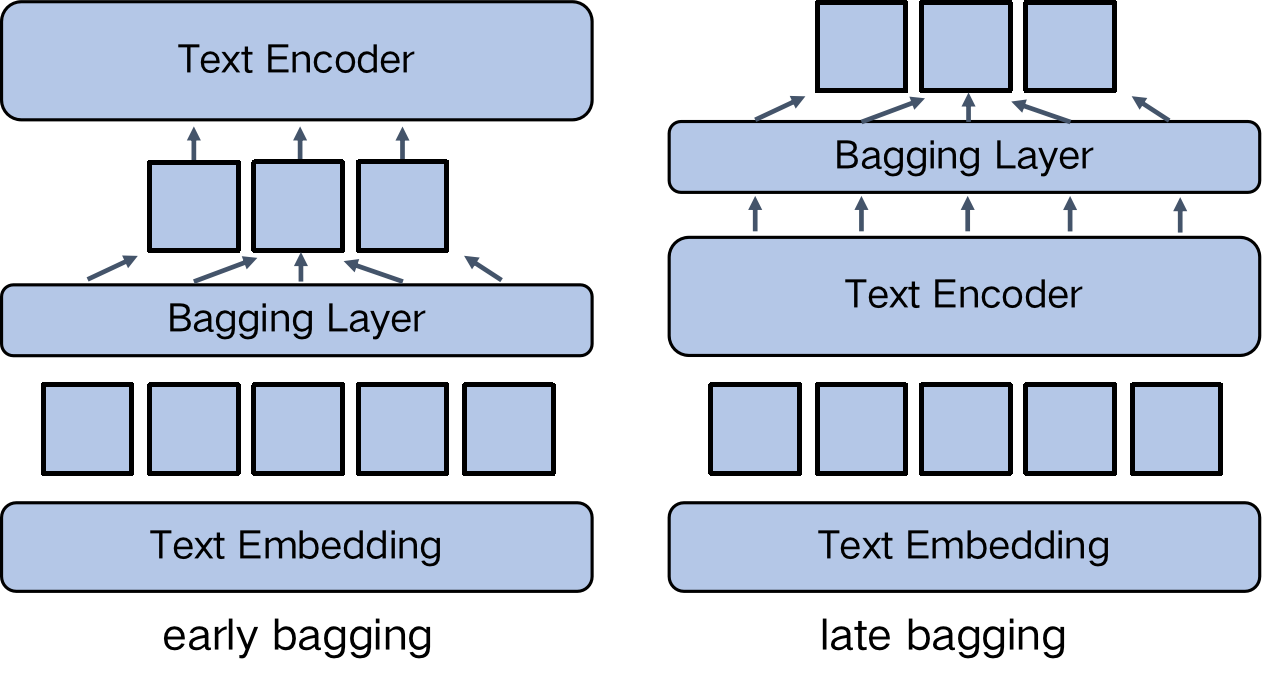}
\caption{Early bagging architecture(left) and late bagging Architecture(right).}
\label{figure:early vs late}
\end{figure}

Two architectures were designed for BagFormer: an early bagging architecture and a late bagging architecture. 
As shown in Figure \ref{figure:early vs late}, the text tokens are bagged before the text encoder in the early bagging architecture. 
For the late bagging architecture, all tokens are bagged after passing through the text encoder.
According to the results, the late bagging architecture outperforms the early bagging architecture, which we will discuss in next section.

\subsection{Implementation Details}
Text encoder is a 6-layer BERT model with 58.3M parameters and image encoder is a 12-layer visual transformer ViT-B/16 with 85.8M parameters. Following \cite{radford2021learning}, we tokenize the text using byte pair encoding (BPE)\cite{sennrich2015neural}.

\begin{table*}[ht]
\centering
\begin{tabular}{ccccccccc}
\toprule[1pt]
Dataset & Method & \multicolumn{3}{c}{Image-to-Text Retrieval} & \multicolumn{3}{c}{Text-to-Image Retrieval} & MR \\
 & & R@1 & R@5 & R@10 & R@1 & R@5 & R@10 & \\
\midrule[1pt]
    \multirow{3}{*}{Flickr30k-CNA} & 
    \textbf{BagFormer(Base)} & \textbf{4.10} 
    & \textbf{15.90} & \textbf{23.20} & \textbf{5.02} & \textbf{13.94} & \textbf{20.62} & \textbf{13.79} \\
    & w/o bagging layer & 1.00 & 5.30 & 8.70 & 4.04 & 12.22 & 18.02 & 8.21 \\
    & w/o late interaction & 0.10 & 0.50 
    & 1.40 & 0.76 & 3.04 & 5.54 & 1.89 \\
\hline
    \multirow{3}{*}{Wukong50k} & \textbf{BagFormer(Base)} & \textbf{10.53} 
    & \textbf{18.50} & \textbf{22.12} & \textbf{18.61} & \textbf{30.06} & \textbf{34.14} & \textbf{22.33} \\
    & w/o bagging layer & 4.31 & 10.91 & 14.68 & 17.46 & 28.60 & 32.75 & 18.12 \\
    & w/o late interaction & 0.07 & 0.18 
    & 0.32 & 11.31 & 19.64 & 23.42 & 9.16 \\
\hline
    \multirow{5}{*}{MUGE} & 
    ALBEF & 30.81 & 57.74 & 67.94 & 43.01 & 68.73 & 77.17 & 57.57 \\
    & Wukong\textsubscript{ViT-B} & - & - & - & 33.40 & 59.30 & 69.70 & - \\
    & \textbf{BagFormer(Base)} & \textbf{31.11} & \textbf{57.65} & \textbf{67.57} & \textbf{40.73} & \textbf{67.61} & \textbf{76.59} & \textbf{56.88} \\
    & w/o bagging layer & 2.52 & 8.91 & 15.48 & 37.50 & 64.39 & 74.04 & 33.81 \\
    & w/o late interaction & 1.08 & 3.69 
    & 5.89 & 19.05 & 39.15 & 49.02 & 19.65 \\
\bottomrule[1pt]
\end{tabular}
\caption{
Ablation study of zero-shot image-text retrieval on several datasets.
}
\label{table:zero-shot}
\end{table*}

\begin{table*}[ht]
\centering
\begin{tabular}{ccccccccc}
\toprule[1pt]
Dataset & Method & \multicolumn{3}{c}{Image-to-Text Retrieval} & \multicolumn{3}{c}{Text-to-Image Retrieval} & MR \\
 & & R@1 & R@5 & R@10 & R@1 & R@5 & R@10 & \\
\midrule[1pt]
    \multirow{3}{*}{Flickr30k-CNA} & 
    \textbf{BagFormer(Base)} & \textbf{66.0} & \textbf{87.1} & \textbf{92.4} & \textbf{48.19} & \textbf{76.75} & \textbf{85.51} & \textbf{75.99} \\
    & w/o bagging layer & 35.6 & 65.6 & 78.6 & 44.35 & 73.89 & 82.97 & 63.50 \\
    & w/o late interaction & 27.4 & 58.5 
    & 72.2 & 30.19 & 59.28 & 70.32 & 52.98 \\
\hline
    \multirow{5}{*}{MUGE} & 
    ALBEF & 41.86 & 72.78 & 82.27 & 53.05 & 80.47 & 87.48 & 69.65 \\
    & Wukong\textsubscript{ViT-B} & - & - & - & 39.20 & 66.90 & 77.40 & - \\
    & \textbf{BagFormer(Base)} & \textbf{41.23} & \textbf{72.06} & \textbf{81.10} & \textbf{52.41} & \textbf{79.19} & \textbf{86.54} & \textbf{68.75} \\
    & w/o bagging layer & 11.34 & 30.14 & 41.18 & 47.20 & 75.69 & 84.42 & 48.33 \\
    & w/o late interaction & 21.29 & 51.16 
    & 65.24 & 46.44 & 75.28 & 84.36 & 57.29 \\
\bottomrule[1pt]
\end{tabular}
\caption{
Ablation study of fine-tuned image-text retrieval on several datasets.
}
\label{table:fine-tuned}
\end{table*}

\section{Experiments}

\subsection{Experimental Setup}
\textbf{Pre-training Dataset}. Our pre-training dataset consists 108M image-text pairs, collected from the web. Average length of text is 25 words.

\textbf{Experiment Details}. 
We pre-train the model for 15 epochs using a batch size of 512 on 8 NVIDIA V100 GPUs. AdamW\cite{loshchilov2017decoupled} optimizer is used with a weight decay of 0.02. Following a cosine schedule, the learning rate is warmed up to $1e^{-4}$ in the first 2000 iterations, then decayed to $1e^{-5}$. Random cropped images of 256 x 256 resolution and RandAugment\cite{cubuk2020randaugment} are used as inputs during pre-training. During fine-tuning, we keep the image resolution at 256 × 256. During inference, we resize the images without cropping them. 
In addition, we constructed a vocabulary of 342,729 entities for the bagging layer.

\subsection{Modal Granularity Mismatch}
Our experiments have observed a discrepancy in performance between text-to-image recall and image-to-text recall in cross-modal retrieval. 
In Table \ref{table:zero-shot} and Table \ref{table:fine-tuned}, we see that the image-to-text retrieval performance of the CLS token model(w/o late interaction) and the tokenwise interaction model(w/o bagging layer) is significantly lower than their text-to-image retrieval performance.
However, the image-to-text retrieval performance of the BagFormer model is significantly better than the control group(model w/o late interaction and model w/o bagging layer). 
We can conclude that the improved performance of the BagFormer model is mainly due to its improved image-to-text retrieval task.

Our findings indicate that the poor performance of the CLS token model (without late interaction) and the tokenwise interaction model (without a bagging layer) is primarily due to a granularity mismatch between the image and the text.
This is especially true in languages such as Chinese, in which token granularity is not sufficient to convey the complete semantics. 
BagFormer attempts to solve this problem by converting token granularity into bags of words or entities to enable better alignment between the text and the images. 
The success of BagFormer is largely attributed to its ability to alleviate the modal granularity mismatch.

\begin{table*}[h]
\centering
\begin{tabular}{ccccccccc}
\toprule[1pt]
Setting & Method & \multicolumn{3}{c}{Image-to-Text Retrieval} & \multicolumn{3}{c}{Text-to-Image Retrieval} & MR \\
 & & R@1 & R@5 & R@10 & R@1 & R@5 & R@10 & \\
 \midrule[1pt]
    \multirow{2}{*}{Zero-shot} & BagFormer\textsubscript{early-bagging} & 19.83 & 34.22 & 38.97 & 40.31 & 67.95 & 76.87 & 46.36 \\
    & BagFormer\textsubscript{bagwise} & 31.11 & 57.65 & 67.57 & 40.73 & 67.61 & 76.59 & 56.88 \\
\hline
    \multirow{2}{*}{Fine-tuned} & BagFormer\textsubscript{early-bagging} & 39.61 & 67.60 & 74.75 & 51.61 & 77.75 & 85.12 & 66.07 \\
    & BagFormer\textsubscript{bagwise} & 41.23 & 72.06 & 81.10 & 52.41 & 79.19 & 86.54 & 68.75 \\
\bottomrule[1pt]
\end{tabular}
\caption{
Results of image-text retrieval for early and late bagging on the MUGE dataset.
}
\label{table:early vs. late}
\end{table*}

\subsection{Image-Text Retrieval}

In this section, we test our models on two subtasks: image-to-text retrieval and text-to-image retrieval. 
In the image-to-text retrieval, the model retrieves a target text from a set of candidates given an image as query, or vice versa for the text-to-image retrieval. 
We evaluate our BagFormer on three retrieval benchmark datasets, including Flickr30K-CNA\cite{xie2022zero}, MUGE\cite{lin2021m6} and Wukong50K\cite{gu2022wukong}, under both zero-shot and fine-tuned settings. Due to the lack of a training set, Wukong50K is used only in zero-shot setting.

Our pre-trained models are evaluated in zero-shot and fine-tuned settings for each dataset.
In accordance with common practices, we report Recall@K (recall of the top K candidates) with K = 1, 5, 10 for both image-to-text retrieval and text-to-image retrieval across all datasets. 
In the final comparison, the mean recall (MR) of Recall@K is used. 
Results are reported for the test sets, except for MUGE, where there are no test set available. 
For that case, we report results from the MUGE validation set.

Table \ref{table:zero-shot} and Table \ref{table:fine-tuned} summarize the results of the zero-shot and fine-tuned image-text retrieval respectively. 
ALBEF is a very strong benchmark for single-tower model, which has the highest results in MUGE dataset. 
The result of our BagFormer achieve comparable performance to state-of-the-art model ALBEF, while significantly higher than another benchmark model Wukong\textsubscript{ViT-B}, across different datasets, in either zero-shot or fine-tuned settings. 
In comparison to ALBEF, our dual tower BagFormer, which has comparable performance.

\subsection{Ablation Study}

Additionally, we conducted ablation studies as shown in table \ref{table:zero-shot} and table \ref{table:fine-tuned}. BagFormer w/o bagging layer remove the bagging layer but keep the late interaction, which is the token-wise model as in ColBERT\cite{khattab2020colbert} and FILIP\cite{yao2021filip}. BagFormer w/o late interaction remove both the bagging layer and late interaction loss, which is similar to CLIP\cite{radford2021learning}. 

As shown in ablation studies, these improvements are owning to the novel designs of bagging layer and late interaction. 
We also notice that bagging layer plays a key role in boosting the retrieval performance. 
The main improvement comes from Image-to-Text retrieval task, which alleviate the modal granularity mismatch.

\subsection{Efficiency Analysis}
The efficiency of BagFormer was evaluated using an NVIDIA T4-16G GPU. Latency and throughput of the model were assessed in a re-ranking setting, in which each query is paired with the top 64 candidates and rescored by the model. 
The results, as presented in Table \ref{table:Efficiency analysis}, demonstrate that BagFormer has a significantly shorter latency (20.72 times shorter) and higher throughput (25.74 times higher) compared to the single encoder architecture, which use ALBEF as the feature extractor and 6-layer transformer to fuse multimodal features. 
Additionally, while BagFormer is slightly slower than the dual encoder architecture, in which ALBEF was used as the feature extractor and the CLS token similarity was employed for scoring, it demonstrates much higher recall metrics.

\begin{table}[h]
\centering
\begin{tabular}{ccc}
\toprule[1pt]
Method & \makecell{Latency \\ (ms)} & \makecell{Throughput \\ (queries/s)} \\ \midrule[1pt]
All-to-all interaction & 840.80 & 1.26 \\ \hline
Representation-based similarity & 40.41 & 29.30 \\ \hline
Bag-wise interaction & 40.57 & 27.03 \\ \bottomrule[1pt]
\end{tabular}
\caption{
Efficiency analysis. 
}
\label{table:Efficiency analysis}
\end{table}

\subsection{Early bagging vs. Late bagging}

The early bagging and late bagging architectures were tested on the MUGE dataset\cite{lin2021m6}. 
Table \ref{table:early vs. late} shows that late bagging architecture generally outperforms early bagging architecture on cross-modal retrieval tasks.
In zero-shot results, the late bagging architecture performed better than the early bagging architecture, indicating that it can help the model to obtain better text-image alignment during pre-training.
Compared with early bagging architecture, the reason why late bagging architecture is better may be that there is less information loss and better use of prior information provided by external knowledge.

\begin{figure*}[h]
\centering
\includegraphics[scale=0.65]{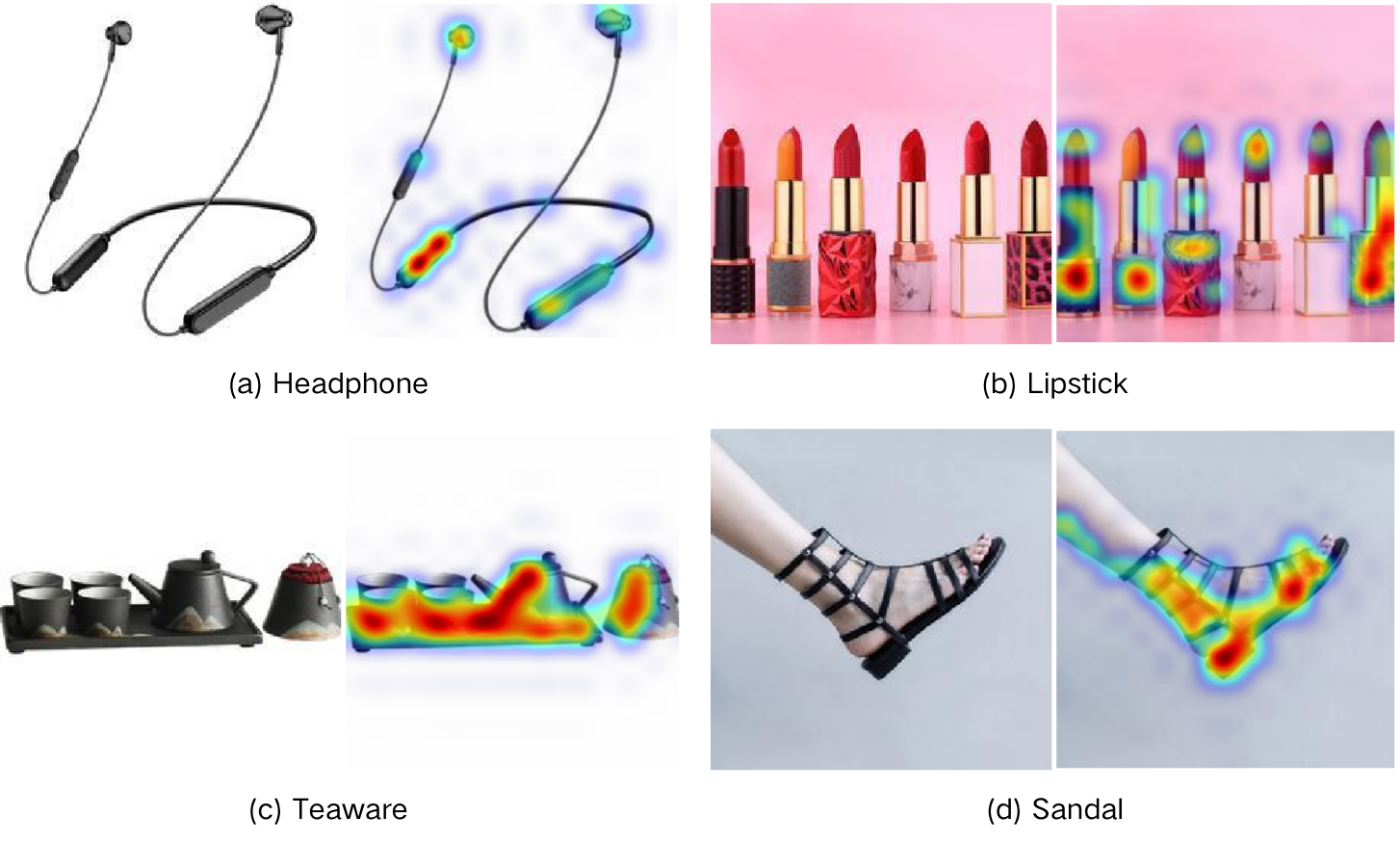}
\caption{Word-patch visualizations on the bag-wise similarity maps.}
\label{figure:visualization}
\end{figure*}

\subsection{Visualization of cross-modal alignment}

The following section examines BagFormer's ability to capture fine-grained cross-modal alignment. Due to a lack of word-patch alignment capability, the visualization of CLIP is excluded\cite{yao2021filip}. As shown in Figure \ref{figure:visualization}, we visualize four images with word-bag in the MUGE dataset. The visualization is performed based on the bag-wise similarity between the image patches and textual bags. Specifically, we calculate the bag-wise similarity between image patches and word bag to produce a heat map that shows the word bag's activation. 

As shown in the Figure \ref{figure:visualization}, BagFormer can learn meaningful fine-grained features and demonstrate promising localization abilities. BagFormer can outline the teaware in the example of Figure \ref{figure:visualization} (c), while BagFormer often aligns to the key part of the target object in the examples of Figure \ref{figure:visualization} (a,b,d). 

\section{Conclusion}
\label{sec:bibtex}
In this paper, we introduce BagFormer, a dual encoder model with novel bag-wise interaction mechanism. 
The proposed BagFormer leverage bag-wise interaction mechanism, which introduce appropriate granularity to text and implicitly introduce entity knowledge. 
Through empirical evaluation on the MUGE, Flickr30k-CNA, and Wukong50k datasets, we demonstrate that BagFormer is capable of achieving performance comparable to that of a single tower model, but with significantly lower latency (20.7 times) and higher throughput (25.74 times). 

\section*{Ethical Statement}

There are no ethical issues.

\bibliographystyle{named}
\bibliography{ijcai22}

\end{document}